\documentclass[twocolumn,superscriptaddress,prl,aps]{revtex4}

\usepackage{graphicx}
\usepackage{dcolumn}
\usepackage{bm}
\usepackage{ulem}

\usepackage{color}

\begin{document}

\title{A Superglass Phase of Interacting Bosons}

\date{\today}

\author{Ka-Ming Tam}
\affiliation{Department of Physics and Astronomy, University of Waterloo, Waterloo, ON, N2L 3G1, Canada}

\author{Scott Geraedts}
\affiliation{Department of Physics and Astronomy, University of Waterloo, Waterloo, ON, N2L 3G1, Canada}

\author{Stephen Inglis}
\affiliation{Department of Physics and Astronomy, University of Waterloo, Waterloo, ON, N2L 3G1, Canada}

\author{Michel J. P. Gingras}
\affiliation{Department of Physics and Astronomy, University of Waterloo, Waterloo, ON, N2L 3G1, Canada}
\affiliation{Canadian Institute for Advanced Research, 180 Dundas St. W., Toronto, ON, M5G 1Z8, Canada}

\author{Roger G. Melko}
\email{rgmelko@uwaterloo.ca}
\affiliation{Department of Physics and Astronomy, University of Waterloo, Waterloo, ON, N2L 3G1, Canada}

\begin{abstract}
We introduce a Bose-Hubbard Hamiltonian with random disordered interactions as a model to study the interplay
 of superfluidity and glassiness in a system of three-dimensional hard-core bosons at half-filling.  
Solving the model using large-scale quantum Monte Carlo simulations, 
we show that these disordered interactions promote a stable superglass phase, 
where superflow and glassy density localization coexist in equilibrium without exhibiting phase separation. 
 The robustness of the superglass phase is underlined by its existence in a replica mean-field calculation 
on the infinite-dimensional Hamiltonian.
\end{abstract}

\maketitle


Despite the simplicity of its constituent atoms, the phase diagram of bulk helium has proven to be compelling and controversial -- especially in 
light of recent observations of supersolid behavior in the crystal phase of Helium-4 ($^4$He) \cite{Kim-Chan1,Kim-Chan2}. 
Several experiments have shown that 
superflow is enhanced by imperfections in the crystal lattice, 
such as roaming defects and interstitial atoms \cite{Balibar}.
Most strikingly, new experiments by Davis and collaborators 
indicate
that the onset of superflow is intimately tied to
relaxation dynamics characteristic of glassy, or amorphous, 
solids \cite{Hunt}. 
The precise relationship between superflow and glassiness,
and the possible existence of a novel {\it superglass} (SuG) state,
is not without controversy in the experimental community \cite{West}. 
Even outside of the $^4$He context, the possible existence of a 
bosonic superglass phase
is of broad theoretical importance -- one that can be 
explored for example in simple models of interacting bosons.
Lattice models have been instrumental in explaining some of the
basic phenomenology of strongly-coupled quantum systems, including that of $^{4}$He.
The prototypical model is the lattice Bose-Hubbard (BH) model \cite{BoseHub}.  
Its extended phase diagram is known to contain
superfluids, Mott-insulating ``crystals'', and supersolid phases, where the latter 
refers to a state with coexisting superfluidity and broken translational symmetry 
in the particle density \cite{Leggett}.

Models of interacting bosons
with disorder have been considered for some time \cite{BoseHub,MaLee} 
typically in the local chemical potential, 
as might be realized in $^4$He absorbed in porous media. 
In these and related models \cite{Gimp} arises either a ``Bose-glass'' (BG) phase, with 
localized disorder in the particle densities but no coexisting superflow, 
or superfluid phases with locally inhomogeneous superflow \cite{Long} -- 
neither of which correspond to a SuG state. 
This can be understood in part by considering the nature of the states 
in the BG close to the Mott lobes which, through  
Anderson localization \cite{localization}, 
essentially form single-particle localized states.
For this reason, the BG  cannot support phase coherence; other types of interactions 
beyond the random local chemical potential are necessary to induce superglassiness.
In this paper we show that a thermodynamic SuG state of bosons can be stabilized via random 
pairwise boson-boson interactions.

It is widely believed 
that disorder (in the form of dislocations, grain boundaries, impurities, etc.) 
 plays a role in the supersolid behavior of $^4$He 
\cite{Balibar}, although the precise nature of this role is controversial.
At the microscopic level, $^{4}$He atoms interact via a deterministic pairwise potential in
the continuum.
If perfect crystal order is avoided by an effective 
quenching \cite{Boninsegni} into a glassy phase at low temperature, even
without explicit randomness in the Hamiltonian, 
both disorder and random frustration can be self-generated from the
deterministic potential, similar to scenarios put forward in discussions of the
structural glass  transition \cite{Targus}.
 Therefore, we propose that an effective theory to describe glassy behavior in
$^{4}$He via this disordered and frustrated 
environment is a BH lattice model with {\it randomly frustrated} boson-boson
interactions.
We note that
the phenomenology of 
random frustration
and glassiness can also be constructed
via a mathematical embodiment of a BH model 
on a random graph \cite{Carleo} -- an approach that leaves 
open the question of whether a
SuG exists for real finite dimensional lattices.

While the relevance of our model to the physics of $^{4}$He ultimately 
relies on microscopic details via which disorder and frustration may be self-generated, 
it is of direct pertinence to collective phenomena in
ultracold atomic gases.  There, disorder may be introduced through 
mechanisms such as speckle potentials or superimposed optical lattices with 
incommensurate periods \cite{Sanpera}.

Accordingly, we introduce the hard-core BH model,
\begin{equation}
H = -\sum_{\langle ij \rangle} V_{ij} (n_i - 1/2)( n_j-1/2) - t\sum_{\langle ij \rangle}\big( b_i^{\dagger}b_j + b_i b^{\dagger}_j \big),
\label{ham}
\end{equation}
where $\langle ij \rangle$ indicates nearest-neighbor lattice sites.
Here, $n_i$ is the density operator of bosons hopping on a three-dimensional (3D) cubic lattice, $b_i$ ($b^{\dagger}_i$) is the annihilation 
(creation) operator.
$V_{ij}$ is a quenched random potential with bimodal distribution given by $P(V_{ij})=[\delta(V_{ij}-V) + \delta(V_{ij}+V)]/2$,
reminiscent of the canonical spin glass (SG) models studied extensively 
\cite{SG_book,RF-RG}.
The hopping ($t$) term is the standard kinetic energy term in the BH  model, acting as a source of quantum fluctuations 
which favor a superfluid phase whenever $V/t$ is sufficiently small.  Eq.~(\ref{ham}) is explicitly defined to maintain half-filling. Below, we use large-scale Stochastic Series Expansion (SSE) quantum Monte Carlo (QMC) simulations 
to demonstrate that these simple random disordered interactions can promote a stable superglass phase in the Hamiltonian (\ref{ham}),
where superflow and glassy density localization coexist in equilibrium without exhibiting phase separation.  
Following that, the existence of this superglass phase is corroborated by a replica mean-field calculation on an infinite-dimensional model.

\begin{figure}[!htbp]
\includegraphics[width=0.97 \linewidth,angle=0,clip]{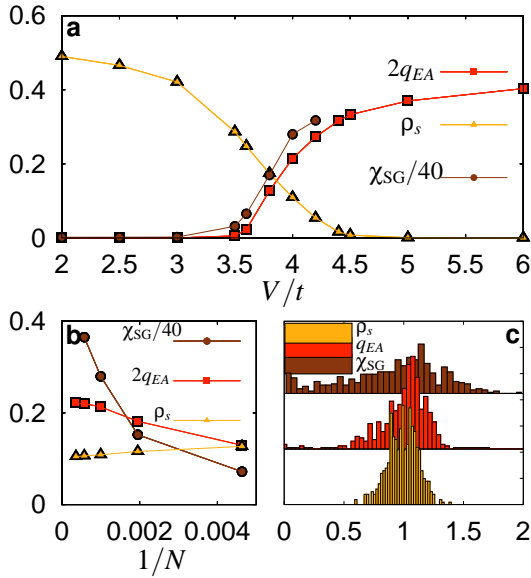}
\caption{  
{\bf a}) The superfluid density, EA order parameter,
and spin glass susceptibility as a function of $V/t$, for $T/t=1/4$ and $L=10$.
For $V/t>4.2$, $\chi_{\rm SG}$ suffers from algorithmic
freezing or loss of ergodicity, making results unreliable without a
refinement of simulation technique (e.g.~parallel tempering).
{\bf b}) Finite-size scaling of the QMC observables in the SuG
state at $V/t=4$ and $T/t=1/4$.
In ({\bf a}) and ({\bf b}), statistical error bars related to QMC sampling are smaller than
the symbol size.
{\bf c}) Histogram of the three observables at $L=10$ in ({\bf b}) using 500 RODs.
The x-axis has been divided by the mean of the distribution in each case, and the y-axis of each has been offset for clarity.
There is no clear separation of the SuG RODs into distinct glassy and superfluid bins.
\label{fig2}
}
\end{figure}

QMC simulations are performed using the finite-temperature SSE technique \cite{SSE},
modified to allow for the random disorder in Eq.~(\ref{ham})
 by employing different directed loop equations for each individual bond interaction. 
A single simulation consists of constructing a 3D cubic lattice (size $N=L^3$) with some unique choice of specific 
bimodal bond distribution, $V_{ij}$, referred to as one realization of disorder (ROD).  Each specific simulation, for a given system size, 
temperature, and interaction strength $V/t$, is equilibrated for $10^6$ MC steps before thermodynamic data is collected (for $2 \times 
10^6$ steps), this MC average denoted below by  $\langle ... \rangle$.  The simulation procedure is repeated for many RODs -- typically 
between $10^2$ and $10^3$, and denoted as $[...]_{\rm avg}$. 

\begin{figure}
\includegraphics[width=0.97\linewidth,angle=0,clip]{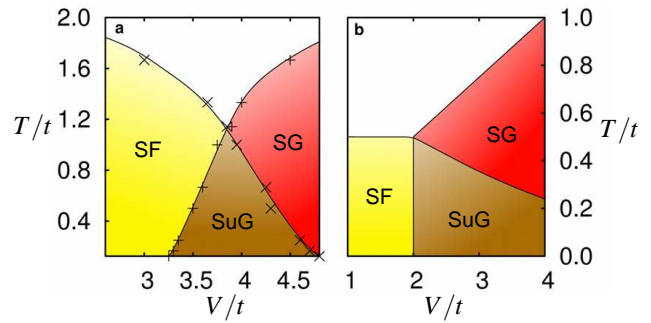}
\caption{ The phase diagram of the ({\bf a}) quantum (Eq.~\ref{ham}) and ({\bf b}) mean-field (Eq.~\ref{Spinham}) models, solved with  
QMC and replica mean-field methods respectively. Both phase diagrams show a robust superglass (SuG) phase, as well as superfluid (SF) 
and quantum spin glass (SG) states.  The data points in {\bf a}, used to determine the phase boundaries, were obtained when $\rho_s$ and $q_{\rm 
EA}$ surpassed an approximate numerical value of $1/\sqrt{N}$ on a $N=12\times 12 \times 12$ system. It is interesting to note that the main 
qualitative difference between the two phase diagrams is the absence of a direct superfluid to superglass transition in the mean-field solution, 
({\bf b}), as a function of temperature.
\label{phase}
}
\end{figure}

We collect data for thermodynamic quantities which are able to quantify both the superfluid and the glassy characteristics of various phases of the 
model. Our definition for superflow is based on the superfluid density,
\begin{equation}
\rho_s = \frac{1}{ 3 \beta N t} \left[{  \sum_{k = x,y,z} \langle W_{k}^2 \rangle }\right]_{\rm avg}
\end{equation}
where $W_k$ is the winding number \cite{Ceperley} measured in the lattice direction $k$. 
Two quantities commonly used in SG models to indicate a glassy phase are the 
Edwards-Anderson (EA) order parameter,
\begin{equation}
q_{\rm EA} =\frac{1}{N} \left[{ \sum_i \langle (n_i-1/2) \rangle^2 }\right]_{\rm avg},
\end{equation}
and the replica overlap parameter \cite{SG_book},
\begin{equation}
q_{ ab} = \frac{1}{N} \sum_i (n_i-1/2)^{a} (n_i-1/2)^{b},
\end{equation}
where two identical (independently simulated) replicas of the system  are labeled  $a$ and $b$ for 
a single ROD.  
The disorder-averaged SG susceptibility is thus defined as
\begin{equation}
\chi_{\rm SG} = N  \left[{  \langle q_{a b}^2 \rangle }\right]_{\rm avg}.
\end{equation}
Figure \ref{fig2} {\bf a} shows these observables as a function of $V/t$. For some finite region 
of $V/t$, the model displays robust superflow (characterized by finite $\rho_s$) {\it and} a finite EA order parameter and SG 
susceptibility -- these trends survive into the thermodynamic limit (see Fig.~\ref{fig2} {\bf b}).  This indicates that the SuG is a stable 
equilibrium phase in this model, persisting down to zero temperature (Fig.~\ref{phase}). 

The simplicity of Hamiltonian (\ref{ham}) suggests a clear mechanism for the formation of the SuG phase; it arises from 
the competition of disorder in the boson interactions with quantum fluctuations. One may therefore wonder whether this mechanism is robust enough  
to be found in standard analytical treatments of well-studied spin glass models.  We address this question by employing an exact 
mapping of the hard-core boson model in Eq.~(\ref{ham}) to a quantum spin model through the transformation to spin-1/2 operators: $S^z_i = 
n_i-1/2$, $S^-_i = b_i$, and $S^+_i = b^{\dagger}_i$.  Equation \ref{ham} can then be written as a standard XXZ model:
\begin{equation}
H = -\sum_{\langle ij \rangle} V_{ij} S^z_i S^z_j - J_{xy} \sum_{\langle ij \rangle}\big( S^x_i S^x_j + S^y_i  S^y_j \big),
\label{Spinham1}
\end{equation}
where $J_{xy}=2t$ and the off-diagonal part of the Hamiltonian is the XY model, which (e.g.~for $V=0$) has
in-plane ferromagnetic long-range order,
$\langle S^x \rangle \neq 0$, corresponding to the superfluid state in the boson language. One now recognizes the diagonal part of (\ref{Spinham1}) 
as the standard EA Ising spin model, 
for which the expected SG phase is reproduced for our model as $t \rightarrow 0$ (Figs~\ref{fig2} and \ref{phase}).

\begin{figure*}
\includegraphics[width=0.8\linewidth,angle=0,clip]{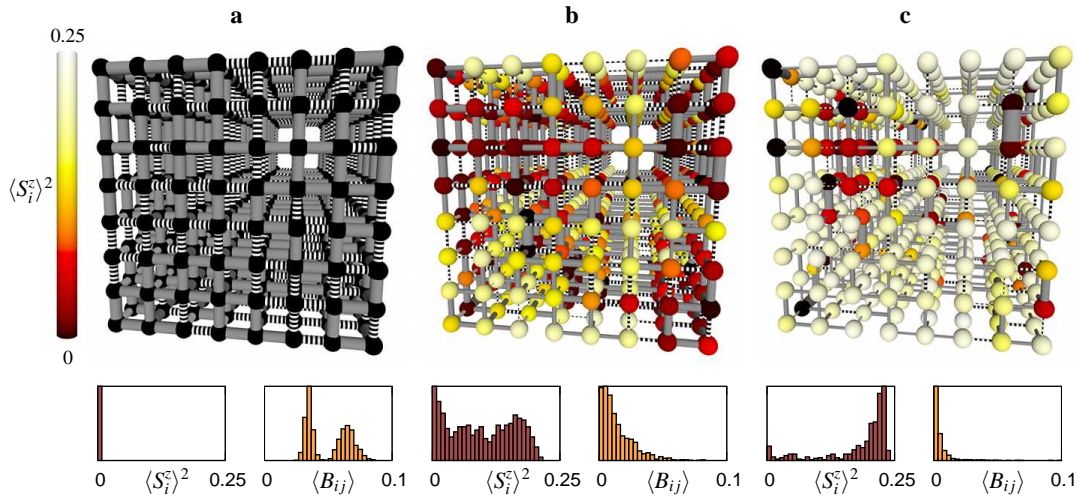}
\caption{ The QMC simulation cell, with one random realization of disorder, averaged over $2 \times 10^6$ Monte Carlo (MC) steps.  Grey bonds 
correspond to $V_{ij}=V$, striped bonds to $V_{ij} = -V$.  On the sites $i$ of the lattice, the density localization is illustrated by the squared 
average of $S^z_i = n_i - 1/2$, which is black for a site where a boson is delocalized (a ``superfluid site''), 
white for a localized (frozen) site, 
and colored for partially localized Boson densities ($0 \leq \langle S^z_i \rangle^2 \leq 0.25$). The bond-strength, defined as the boson hopping 
operator $\langle B_{ij}\rangle = \langle b_i^{\dagger}b_j + b_i b^{\dagger}_j \rangle$ is represented as the thickness of the bonds.  At bottom, 
the histograms of $\langle S^z_i\rangle^2$ (left), and $\langle B_{ij}\rangle$ (right) for each case: superfluid ({\bf a}) at $V/t=3$, 
superglass ({\bf b}) at $V/t=4$, and spin glass ({\bf c}) at $V/t=5$.
The double-peaked $\langle B_{ij}\rangle$ histogram in ({\bf a}) simply results from the bimodal distribution of $V_{ij}$.
\label{fig1}
}
\end{figure*}

An analytical solution for the 3D EA model ($t=0$) is not available. However, 
the mean-field (MF) Sherrington-Kirkpatrick model can be solved  exactly by Parisi's
replica symmetry breaking (RSB) ansatz \cite{Parisi_RSB1}. 
Here, we generalize the replica MF method to include the effects from both the conventional 
$V_{ij}$ Ising coupling, and the off-diagonal  $J_{xy}$-coupling
 \cite{Bray-Moore}.
The MF model is
\begin{equation}
H_{MF} = -\sum_{ij} V_{ij} S^z_i S^z_j - \frac{J_{xy}}{N} \sum_{ij}\big( S^x_i S^x_j + S^y_i  S^y_j \big) \label{Spinham},
\end{equation}
where the summation is over all distinct pairs of spins. 
For simplicity, we use a Gaussian distribution, 
$P(V_{ij}) = \frac{1}{\sqrt{ 2 \pi V^{2} / N}} \exp \Big( \frac{-N V_{ij}^{2}}{ 2 V^{2}} \Big)$, which 
does not qualitatively change the physics compared to a bimodal $P(V_{ij})$.
The disorder averaged free energy using the replica formalism is given by $\beta N f = -[\ln Z]_{\rm avg} = - \lim_{n\rightarrow0} \frac{1}{n} 
([Z^n]_{\rm avg}-1)$, where $[...]_{\rm avg}$ denotes disorder averaging. We introduce the Hubbard-Stratonovich (HS) fields for SG overlap $Q^{ab} 
= \langle S_{z}^{a}S_{z}^{b} \rangle~\forall~a \neq b$, overlap within the same replica $R^{a} =  \langle S_{z}^{a}S_{z}^{a}\rangle$, and off-diagonal 
ordering parameter  
$M^{a} = \langle S_{x}^{a}\rangle$. We assume that $M^a = M$ and $R^a = R$ are replica independent and, for simplicity, consider 
solely a 1-step RSB for $Q^{ab}$. Following Parisi's parameterization for 1-step RSB \cite{SG_book}, we divide the $n$ replicas
 into $n/m$ groups of $m$ 
replicas, and set $Q_{ab} = Q_{1}$ if $a$ and $b$ belong to the same group; or $Q_{ab} = Q_{0}$  if $a$ and $b$ belong to different 
groups. 
Assuming that the HS fields are static, we obtain the free energy at 1-step RSB, 
\begin{eqnarray}
\beta f = - \frac{1}{4} (\beta V)^2 \Big[ (R-Q_1)(1/2-R-Q_1)  \nonumber \\
          - m (Q_{1}^{2}-Q_{0}^{2}) \Big] + \frac{1}{2} \beta J_{xy} M^{2} \nonumber \\
          - \frac{1}{m} \int Dx \ln \left[{ 2 \int Dy \cosh^m(\beta/2 \sqrt{H_{z}^{2}+(J_{xy}M)^2}) }\right],
\label{RSB_free_energy}
\end{eqnarray}
where $\int Dx \hspace{1mm} F(x) \equiv (1/\sqrt{2\pi}) \int_{-\infty}^{\infty} dx \exp(-x^2/2) F(x)$ , and $H_z = V(x\sqrt{Q_{0}}
+ y\sqrt{Q_{1}-Q_{0}})$. The replica symmetric free energy can be recovered by setting any one of the conditions $Q_{1}=Q_{0}$, $m=0$ or $m=1$. 
The self-consistent MF equations are obtained by optimizing the free energy. We find four phases analogous to those 
in the hard-core boson model (see Fig.~\ref{phase}).  In particular, a SuG phase, defined by coexisting in-plane order $M\neq 0$ (the magnetic 
analog to the superfluid order parameter), and SG overlap parameters $Q_{0}\neq0$ and $Q_{1}\neq0$ is clearly obtained.
The vertical nature of the SuG-SF phase boundary originates from
the observation that the replica theory does not admit a self-consistent set of 
equations with nonzero $M$ and $Q_0$ when $V/t \leq 1$.
Physically, this results from the absence of phase fluctuations 
in the superfluid phase when the SG order parameter $Q_0$ is nonzero.

In conclusion, both large-scale QMC simulations and replica MF calculations exhibit a robust,
 stable, equilibrium superglass phase in the 
BH  model, Eq.~(\ref{ham}). The simplicity of the model reveals that the SuG phase 
develops under competition between random disordered 
interactions and quantum fluctuations. QMC simulations allow us to characterize
 the SuG phase in great detail; for example, as evident from 
the single ROD illustrated in Fig.~\ref{fig1}, there is no phase separation 
into distinct ``super'' and ``glassy'' regions within the simulation 
cell. Rather, the bulk of the particles in the system exhibit both off-diagonal
 superfluid and diagonal glassy ordering. 
Most importantly,
our results clearly 
indicate that glassiness can co-exist with superfluidity in a general bosonic 
system in three dimensions. 
Although we did not directly study a microscopic model of $^{4}$He, 
in the event that a metastable SuG \cite{Boninsegni} does exist in 
$^{4}$He within experimentally-accessible timescales \cite{Hunt}, 
quantum fluctuations and random frustration are the most relevant, 
if not inevitable, physical ingredients at play.
Our model provides an affirmative 
answer that a SuG is possible under these two conditions. 
Furthermore,
 since the feasibility of obtaining 
disordered interactions similar to our $V$ term [Eq.~(\ref{ham})]
 has recently been demonstrated in Fermi-Bose mixtures in 
random optical lattices \cite{Sanpera},
such cold atomic systems might also be suitable places to look for superglass phases in 
the future. 
With recent advances in both cold atom experiments and numerical methods \cite{QMCising},
the broad class of disordered Hamiltonian introduced in this paper provides new experimental and theoretical motivation to study the interplay between superfluidity and quantum glassiness in numerous extensions of our model.

\bibliography{superglass_u.bbl}

\begin{thebibliography}{33}
\bibitem[{RF-()}]{RF-RG}
\bibinfo{note}{The random positive and negative signs chosen for $V_{ij}$ are
  for computational convenience. Even if the bare $V_{ij}$ were strictly
  positive, in presence of randomness, positive and negative effective $V_{ij}$
  would be generated under renormalization.}
\bibitem[{QMC()}]{QMCising}
\bibinfo{note}{We note that the quantum fluctuation term in
  Eqs.~(\ref{ham},\ref{Spinham1}) significantly helps QMC simulations to
  equilibriate at temperatures far below those normally accessible to
  simulations of otherwise classical Ising spin glass models.}
\end{thebibliography}

\subsection{Acknowledgments}
We thank M. Lawler and S. Haas for enlightening discussions.
This work has been supported by NSERC, 
CIFAR, and the CRC program (M.J.P.G. Tier 1).
Simulations were made possible by the facilities of SHARCNET.

\end{document}